\documentclass[onecolumn,showpacs,showkeys,preprintnumbers,amsmath,amssymb]{revtex4}

\usepackage{graphicx}% Include figure files
\usepackage{dcolumn}% Align table columns on decimal point
\usepackage{bm}% bold math

\newcommand{\ket}[1]{\ensuremath{\left| #1 \right \rangle}}
\DeclareMathOperator{\asin}{asin}

\begin{document}

\preprint{ }

\title{Transformation of quantum states using uniformly controlled rotations}

\author{Mikko M\"ott\"onen}
\email{mpmotton@focus.hut.fi}
\author{Juha J. Vartiainen}
\author{Ville Bergholm}
\author{Martti\ M.\ Salomaa}

\affiliation{%
Materials Physics Laboratory, P.O. Box 2200, FIN-02015
Helsinki University of Technology, Finland\\
}%

% tekemistä: uniformly n-fold controlled vs. n-fold uniformly controlled

\date{\today}

\begin{abstract}
We consider a unitary transformation which maps any given state of an
$n$-qubit quantum register into another one. This transformation
has applications in the initialization of a quantum
computer, and also in some quantum algorithms.
Employing uniformly controlled rotations, we present a quantum circuit
of $2^{n+2}-4n-4$ CNOT gates and $2^{n+2}-5$ one-qubit elementary
rotations that effects the state transformation.
The complexity of the circuit is noticeably lower than the previously
published results.
Moreover, we present an analytic expression for the rotation angles
needed for the transformation.
\end{abstract}


\maketitle

\section{Introduction}

Quantum algorithms are based on unitary transformations and
projective measurements acting on a quantum register of $n$ qubits~\cite{NielsenChuang}.
Successful execution of an algorithm usually requires a certain initial state as input.
However, depending on the physical realization of the quantum computer,
available initialization procedures may only produce a limited range
of states which may not contain the desired initial state.
This brings up the problem of state preparation, i.e., how to
implement the transformation of an arbitrary quantum state into another one.

The recent progress~\cite{PRL,mottonen,Shende_matrix} in implementing general $n$-qubit gates
using elementary gates has resulted in efficient gate synthesis techniques including
uniformly controlled rotations~\cite{mottonen},
and more recently, quantum multiplexors~\cite{Shende_matrix}.
These techniques are amenable also for implementing quantum gates of certain special classes
of unitary transformations, such as
incompletely specified transformations.
These transformations have been reacently discussed in Ref.~\cite{Shende_vector}, in which an efficient gate decomposition was given in the case of two qubits.

The complexity of a quantum circuit is measured by the number of
elementary gates included. Generally, elementary gates are unitary
transformations acting on one or two qubits. We take the library of
elementary gates to be the conventional set of the controlled NOT
(CNOT) gate and all one-parameter rotations acting on a single qubit.
We omit the phase gate since the
global phase of the state vector has no physical meaning.

The configuration space of the $n$-qubit quantum register is
$2^n$-dimensional complex space. Excluding the global phase and state normalization,
we find that the general unitary transformation transforming a given $n$-qubit state into another
must have at least $2 \times 2^n - 2$ real degrees of freedom. Hence,
in the worst-case scenario, the corresponding quantum circuit should
involve at least $2^{n+1} - 2$ elementary rotations, each carrying
one degree of freedom.
Since each of the CNOT gates can bind at most four elementary
rotations~\cite{shende}, at least
$\lceil \frac{1}{4} (2^{n+1} - 3n - 2) \rceil$ of them
are needed. However, no quantum circuit construction embodying the
minimal complexity has been presented in the literature.
Previously, the upper bound for the number of gates needed for state
preparation has been considered by Knill~\cite{knill}, who found that
no more than $O(n2^n)$ gates are needed for the circuit implementing
the transformation. More recently, a sufficient circuit of $O(2^n)$ elementary gates was obtained as a special case of the
method developed for QR decomposition of a general quantum gate in Ref.~\cite{PRL}, which was also pointed out in Ref.~\cite{Shende_vector}.

In this paper, we describe in detail how to build a quantum circuit
for making a given quantum state transformation employing the uniformly
controlled rotations.
We begin from the transformation which equalizes the phases of the
elements of the input vector $\ket{a}$ and rotates it
to the direction of the basis vector $\ket{e_1}$. In the next phase
the absolute values of elements of the target vector $\ket{b}$
are generated and finally
the phases are adjusted to match of those of $\ket{b}$.
We simplify the circuit by merging certain consecutive gates together.
The resulting quantum circuit of $2^{n+2}-4n-4$ CNOT
gates and $2^{n+2}-5$ one-qubit elementary rotations gives, in
principle, the exact transformation from an $n$-qubit quantum state
$\ket{a}$ into the desired one $\ket{b}$.

\section{Uniformly controlled rotations}

The quantum state of an $n$-qubit register may be described by a complex vector of the form
\begin{equation}\label{tilavektori}
\ket{a} =
\begin{pmatrix}
a_1 \\ a_2 \\ \vdots \\ a_{N}
\end{pmatrix}
=\sum_{i=0}^{N-1} a_{i+1} \ket{b_1^i b_2^i \ldots b_N^i},
\end{equation}
where $N=2^n$, $b_j$ denotes the state of the $j^{\rm th}$ qubit, and
the bit string $b_1^ib_2^i\ldots b_N^i$ is the binary presentation of
the integer~$i$. The state is taken to be normalized to
unity. Furthermore, the overall phase of the state is not observable
and thus irrelevant. This means that an $n$-qubit state has
$2^{n+1} - 2$ real degrees of freedom.
Quantum gates are linear transformations on the space of these vectors and,
hence, may be represented by $N\times N$ unitary matrices.

\begin{figure}
\begin{center}
\includegraphics[width=10cm]{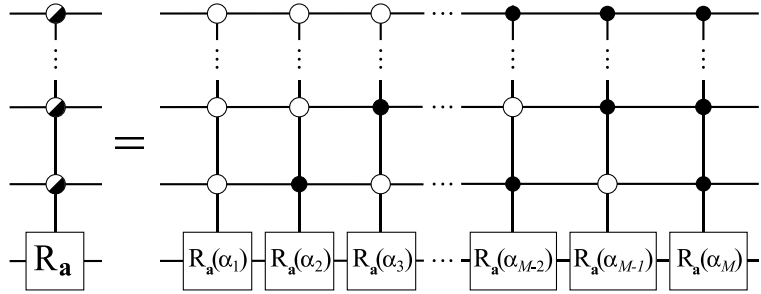}
\end{center}
\caption{Definition of the $k$-fold uniformly controlled rotation
$F^k_m({\bf a},\bm{\alpha})$ of qubit~$m$ about the axis ${\bf a}$.
The left hand side defines the gate symbol used for the uniformly
controlled rotation. The enumeration of the qubits is arbitrary with
the exception that the target qubit is the $m^{\rm th}$ one. The black
control bits stand for value $1$ and the white for $0$. Above,
$M=2^{k}$.
\label{fig1}}
\end{figure}

A uniformly controlled rotation $F^k_m({\bf a},\bm{\alpha})$ is a
quantum gate defined by the $k$ controlled qubits,
the target qubit~$m$, the rotation axis ${\bf a}$ and the angles
$\{\alpha_i\}$, see Ref.~\cite{mottonen}.
As shown in Fig.~\ref{fig1}, the uniformly controlled rotation
corresponds to a sequence of controlled $R_{{\bf a}}({\alpha_i})$
rotations, which covers all the $2^k$ possible control bit sequences. Here
\begin{equation}\label{ra}
R_{{\bf a}}(\alpha_i) = e^{i  {\bf a} \cdot \bm{\sigma}\alpha_i/2} =
I_{2\times 2} \cos \frac{\alpha_i}{2} + i \left( {\bf a} \cdot \bm{\sigma}
\right)\sin \frac{\alpha_i}{2},
\end{equation}
where $\sigma_x$, $\sigma_y$, and $\sigma_z$ are the Pauli matrices and
the dot product ${\bf a} \cdot\bm{\sigma}=\sigma_xa_x+\sigma_ya_y+\sigma_za_z$.

\begin{figure}
\begin{center}
\includegraphics[width=10cm]{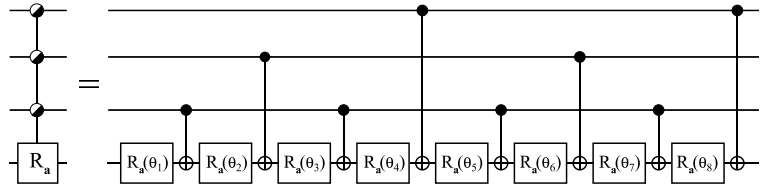}
\end{center}
\caption{\label{fig2} Efficient gate decomposition for the uniformly
controlled rotation $F^3_4({\bf a},\bm{\alpha})$. The relation of the
angles $\{\theta_j\}$ to the angles $\{\alpha_j\}$ is shown in
Eq.~(\ref{linear}).}
\end{figure}

Figure~\ref{fig2} reviews a construction for $F^k_m({\bf a},\bm{\alpha})$
consisting of $2^k$ CNOT gates and $2^k$ one-qubit ${\bf a}$-rotations.
The case $k=3$ is shown.
In the case of a general $k$, the gate sequence may be constructed from the sequence for $k-1$ by
replacing the position of the control in the rightmost CNOT gate
to the new controlled qubit and repeating the obtained sequence twice for suitable rotation angles $\{\theta_j\}$.
The operational principle of the gate sequence requires that $a_x=0$. However, this limitation
can be circumvented by introducing one-qubit basis changing gates on the both side of the gate.

The angles $\{\theta_i\}$ can be obtained from $\{\alpha_i\}$ using
the equation
\begin{equation}
\begin{pmatrix}
\theta_1 \\
\vdots \\
\theta_{2^k}
\end{pmatrix}
= M
\begin{pmatrix}
\alpha_1 \\
\vdots \\
\alpha_{2^k}
\end{pmatrix}, \quad \quad M_{ij} = 2^{-k} (-1)^{b_{j-1} \cdot g_{i-1}},
\label{linear}
\end{equation}
where $b_m$ and $g_m$ stand for the binary code and binary reflected Gray code
representations of the integer $m$, respectively.
In actuality, the position of the controls of the
CNOT gates in Fig.~\ref{fig2} may be chosen in many different ways
which results in replacing $g_{j-1}$ in Eq.~(\ref{linear})
by another cyclic Gray code~\cite{savage:1997}.
Additionally, a horizontally mirrored version of the gate sequence
in Fig.~\ref{fig2} also qualifies to simulate the uniformly
controlled rotation.

\section{State preparation}

We are looking for a gate sequence corresponding
to a matrix $U$ such that $U \ket{a} = \ket{b}$ for given
vectors $\ket{a}$ and $\ket{b}$.
The problem may be reduced to the problem of finding a matrix $V$
which takes an arbitrary vector to some fixed vector $\ket{r}$,
since then we may take $A$ and $B$ such that
$A \ket{a} = \ket{r} = B \ket{b}$ and, hence,
$B^\dagger A \ket{a} = \ket{b}$, where the dagger denotes
the Hermitian conjugate. For convenience, we take the fixed
vector to be the first basis vector
$\ket{e_1} = \ket{00 \ldots 0} = (1,0,0,\ldots,0)^T$.

Our algorithm for transforming
$\ket{a} = (|a_1|e^{i\omega_1},|a_2|e^{i\omega_2},\ldots,|a_N|e^{i\omega_N})^T$
% defined in Eq.~(\ref{tilavektori})
into $\ket{e_1}$ works as follows:
\begin{itemize}
\item
First we equalize the phases $\omega_i$ using a cascade of uniformly
controlled $z$-rotations $\Xi_z$, rendering the vector real up to the global
phase $\phi$: $\Xi_z \ket{a} = e^{i \phi} \ket{\hat{a}}$.
\item
Then we rotate the real state vector $\ket{\hat{a}}$ into the direction of $\ket{e_1}$
using a similar sequence of uniformly controlled
$y$-rotations $\Xi_y$, thus achieving our goal.
\end{itemize}

The first step can be readily accomplished using a general diagonal
$n$-qubit quantum gate first considered in Ref.~\cite{bullock:2004}.
It is efficiently produced by a sequence of uniformly controlled
$z$-rotations as
\begin{equation}
\Xi_z =
\prod_{j=1}^{n} F^{j-1}_{j}({\bf z},\bm{\alpha}^z_{n-j+1})
\otimes I_{2^{n-j}},
\end{equation}
where the gate $F^{j-1}_{j}({\bf z},\bm{\alpha}^z_{n-j+1})$
equalizes the phases of the elements connected through the qubit~$j$. The rotation angles $\{\alpha_{j,k}^z\}_j$,
the elements of $\bm{\alpha}^z_{k}$, are found to be
\begin{equation}
\alpha_{j,k}^z = \sum_{l=1}^{2^{k-1}}(\omega_{(2j-1)2^{k-1}+l}-\omega_{(2j-2)2^{k-1}+l})/2^{k-1} \\\label{alp2},
\end{equation}%
where $j=1,2,\ldots,2^{n-k}$ and $k=1,2,\ldots,n$.

Next we apply a uniformly controlled $y$-rotation
$F_{n}^{n-1}({\bf y},\bm{\alpha}^y)$
with angles
$\{\alpha_j^y\} = \{2\asin \left(
|a_{2j}|/\sqrt{|a_{2j-1}|^2+|a_{2j}|^2} \right) \}$.
This has the effect of zeroing the elements of the vector that correspond
to the states standing for bit value one in the qubit~$n$:
\begin{equation}\label{pieni_kampi}
F_{n}^{n-1}({\bf y},\bm{\alpha}^y) \ket{\hat{a}} =
(a_{1,2},0,a_{2,2},0,\ldots,a_{N/2,2},0)^T =
(a_{1,2},a_{2,2},\ldots,a_{N/2,2})^T \otimes (1,0)^T
\end{equation}
where $\{a_{j,2}\} = \{ \sqrt{|a_{2j-1}|^2 + |a_{2j}|^2} \}$.
In effect we have zeroed the last qubit of the register.
This procedure can be repeated on the remaining nonzero elements,
until we reach $\ket{e_1}=(1,0)^T \otimes \ldots \otimes (1,0)^T$.

\begin{figure}
\begin{center}
\includegraphics[width=14cm]{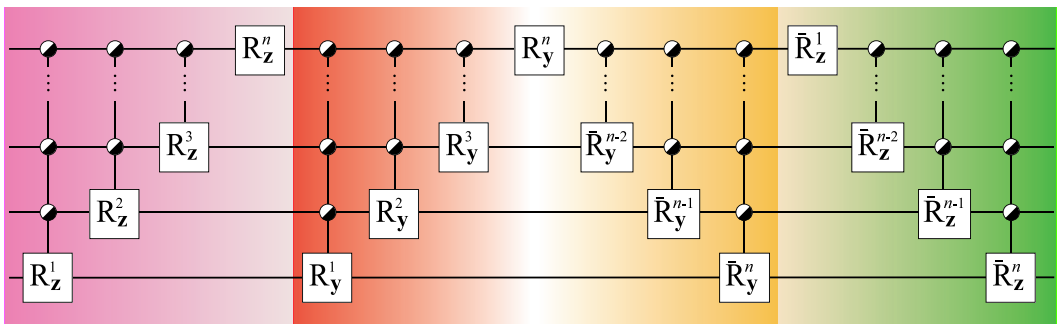}
\end{center}
\caption{\label{fig3} Gate sequence for state preparation using uniformly controlled rotations.
%A shorthanded notation ${R}_{{\bf q}_k}$ has been introduced for the rotation matrices $R_{\bf q}(\alpha_{j,k}^q)$.
The rotation angles $\{\alpha_{j,k}^q\}$ for the uniformly controlled rotations are given in Eqs.~(\ref{alp}) and (\ref{alp2}).}
\end{figure}

Employing the above steps one obtains the desired decomposition
\begin{equation}\label{pre}
\Xi_y \Xi_z \ket{a} =
\left( \prod_{j=1}^{n} F^{j-1}_{j}({\bf y},\bm{\alpha}^y_{n-j+1})
\otimes I_{2^{n-j}} \right)
\left( \prod_{j=1}^{n} F^{j-1}_{j}({\bf z},\bm{\alpha}^z_{n-j+1})
\otimes I_{2^{n-j}} \right)
\ket{a} = e^{i \sum_{j=1}^N \omega_j /N} \ket{e_1}.
\end{equation}
The product of non-commuting matrices in Eq.~(\ref{pre}) is
to be taken from left to right. Here, to eliminate the
remaining global phase one could apply a phase gate.
After solving the recursion, the rotation angles in Eq.~(\ref{pre})
are found to acquire the values
\begin{eqnarray}\label{alp}
\alpha_{j,k}^y & = &2{\rm asin}\left(\sqrt{\sum_{l=1}^{2^{k-1}}|a_{(2j-1)2^{k-1}+l}|^2}/\sqrt{\sum_{l=1}^{2^{k}}|a_{(j-1)2^{k}+l}|^2}\right),
\end{eqnarray}
where $j=1,2,\ldots,2^{n-k}$ and $k=1,2,\ldots,n$.
Fig.~\ref{fig3} shows the quantum circuit corresponding to
Eq.~(\ref{pre}). The resulting gate sequence is slightly simplified by noting that
uniformly controlled $z$-rotations, being diagonal, can always be
commuted through the control bits of another uniformly controlled
gate. Hence, uniformly controlled $z$ and $y$ rotations acting on the same set of qubits can be
commuted next to each other, whereby we can cancel one CNOT from each
gate by mirroring the $y$ gate.

To transform the state $\ket{a}$ to $\ket{b}$ we need to construct
two circuits; the first one takes $\ket{a}$ to $\ket{e_1}$
and the second one  $\ket{e_1}$ to $\ket{b}$.
Since the uniformly $k$-fold controlled rotation may be constructed
from $2^k$ CNOT gates and $2^k$ one-qubit rotations, the
entire state preparation circuit requires
%$4 \cdot (2^{n}-n-1)$ CNOT gates and $4 \cdot (2^{n}-1) - 1$
$2^{n+2}-4n-4$ CNOT gates and $2^{n+2} - 5$
one-qubit rotations.

\section{Conclusion}

In conclusions, we have shown how to construct a general state
preparation circuit using a sequence of uniformly controlled rotations.
The resulting gate sequence of $2^{n+2}-4n-4$
CNOT gates and $2^{n+2}-5$ one-qubit elementary rotations establishes a new
upper bound for the complexity of the transformation.
By counting the degrees of freedom of the problem, we find  a lower bound of $2^{n+1}-2$ for the number of
one-qubit elementary rotations. This implies the lower bound $\lceil \frac{1}{4} (2^{n+1} - 3n - 2) \rceil$ for the number of CNOT gates.

Provided that
the initial or final state coincides with some basis vector $\ket{e_i}$ 
only half of the CNOT and one-qubit rotation gates are needed.
In other special cases some simplifications to the gate sequence also occur.
We have also introduced a closed-form scheme for determining the rotation angles
in such way that an arbitrary state of the quantum register
transforms into desired state. 

The gate count is small compared to the incomplete QR decomposition which takes
 approximately $6.3 \times 2^n$ CNOT gates to transform $\ket{a} \to \ket{e_1}$
and thus $12.6 \times 2^n$ for the whole transformation.
It is still an open question if the transformation could be done more directly, i.e., merging some of the consecutive gates together and finding efficient gate array for implementing them. This would reduce the number of elementary gates needed.

\begin{acknowledgments}
This research is supported by the Academy of Finland through the project ``Quantum Computation'' (No. 206457).
MM thanks the Foundation of Technology (Finland), JJV the Nokia Foundation, MM and VB the Finnish Cultural
Foundation, and MMS the Japan Society for the Promotion of Science for financial support.
Sami~Virtanen is acknowledged for stimulating discussions.
\end{acknowledgments}

\bibliography{stateprep}

\begin{thebibliography}{9}
\expandafter\ifx\csname natexlab\endcsname\relax\def\natexlab#1{#1}\fi
\expandafter\ifx\csname bibnamefont\endcsname\relax
  \def\bibnamefont#1{#1}\fi
\expandafter\ifx\csname bibfnamefont\endcsname\relax
  \def\bibfnamefont#1{#1}\fi
\expandafter\ifx\csname citenamefont\endcsname\relax
  \def\citenamefont#1{#1}\fi
\expandafter\ifx\csname url\endcsname\relax
  \def\url#1{\texttt{#1}}\fi
\expandafter\ifx\csname urlprefix\endcsname\relax\def\urlprefix{URL }\fi
\providecommand{\bibinfo}[2]{#2}
\providecommand{\eprint}[2][]{\url{#2}}

\bibitem[{\citenamefont{Nielsen and Chuang}(2000)}]{NielsenChuang}
\bibinfo{author}{\bibfnamefont{M.~A.} \bibnamefont{Nielsen}} \bibnamefont{and}
  \bibinfo{author}{\bibfnamefont{I.~L.} \bibnamefont{Chuang}},
  \emph{\bibinfo{title}{Quantum Computation and Quantum Information}}
  (\bibinfo{publisher}{Cambridge University Press}, \bibinfo{year}{2000}).

\bibitem[{\citenamefont{Vartiainen et~al.}(2004)\citenamefont{Vartiainen,
  M\"ott\"onen, and Salomaa}}]{PRL}
\bibinfo{author}{\bibfnamefont{J.~J.} \bibnamefont{Vartiainen}},
  \bibinfo{author}{\bibfnamefont{M.}~\bibnamefont{M\"ott\"onen}},
  \bibnamefont{and} \bibinfo{author}{\bibfnamefont{M.~M.}
  \bibnamefont{Salomaa}} (\bibinfo{year}{2004}), \eprint{quant-ph/0312218}.

\bibitem[{\citenamefont{Mikko~M\"ott\"onen
  et~al.}(2004)\citenamefont{Mikko~M\"ott\"onen, Vartiainen, Bergholm, and
  Salomaa}}]{mottonen}
\bibinfo{author}{\bibfnamefont{M.}~\bibnamefont{Mikko~M\"ott\"onen}},
  \bibinfo{author}{\bibfnamefont{J.~J.} \bibnamefont{Vartiainen}},
  \bibinfo{author}{\bibfnamefont{V.}~\bibnamefont{Bergholm}}, \bibnamefont{and}
  \bibinfo{author}{\bibfnamefont{M.~M.} \bibnamefont{Salomaa}}
  (\bibinfo{year}{2004}), \eprint{quant-ph/0404089}.

\bibitem[{\citenamefont{Shende et~al.}(2004)\citenamefont{Shende, Bullock, and
  Markov}}]{Shende_matrix}
\bibinfo{author}{\bibfnamefont{V.~V.} \bibnamefont{Shende}},
  \bibinfo{author}{\bibfnamefont{S.~S.} \bibnamefont{Bullock}},
  \bibnamefont{and} \bibinfo{author}{\bibfnamefont{I.~L.} \bibnamefont{Markov}}
  (\bibinfo{year}{2004}), \eprint{quant-ph/0406176}.

\bibitem[{\citenamefont{Shende and Markov}(2004)}]{Shende_vector}
\bibinfo{author}{\bibfnamefont{V.~V.} \bibnamefont{Shende}} \bibnamefont{and}
  \bibinfo{author}{\bibfnamefont{I.~L.} \bibnamefont{Markov}}
  (\bibinfo{year}{2004}), \eprint{quant-ph/0401162}.

\bibitem[{\citenamefont{Shende et~al.}(2003)\citenamefont{Shende, Markov, and
  Bullock}}]{shende}
\bibinfo{author}{\bibfnamefont{V.~V.} \bibnamefont{Shende}},
  \bibinfo{author}{\bibfnamefont{I.~L.} \bibnamefont{Markov}},
  \bibnamefont{and} \bibinfo{author}{\bibfnamefont{S.~S.}
  \bibnamefont{Bullock}} (\bibinfo{year}{2003}), \eprint{quant-ph/0308033}.

\bibitem[{\citenamefont{Knill}(1995)}]{knill}
\bibinfo{author}{\bibfnamefont{E.}~\bibnamefont{Knill}} (\bibinfo{year}{1995}),
  \eprint{quant-ph/9508006}.

\bibitem[{\citenamefont{Savage}(1997)}]{savage:1997}
\bibinfo{author}{\bibfnamefont{C.}~\bibnamefont{Savage}},
  \bibinfo{journal}{SIAM Rev.} \textbf{\bibinfo{volume}{39}},
  \bibinfo{pages}{605} (\bibinfo{year}{1997}).

\bibitem[{\citenamefont{Bullock and Markov}(2004)}]{bullock:2004}
\bibinfo{author}{\bibfnamefont{S.~S.} \bibnamefont{Bullock}} \bibnamefont{and}
  \bibinfo{author}{\bibfnamefont{I.~L.} \bibnamefont{Markov}},
  \bibinfo{journal}{Quant. Inf. and Comp.} \textbf{\bibinfo{volume}{4}},
  \bibinfo{pages}{27} (\bibinfo{year}{2004}).

\end{thebibliography}

\end{document}